# Nanodiamond sensing in dynamic environments with fast-tracking through four-point positioning

*Guoli Zhu [1] [‡], Ming-Zhong Ai [1] [‡], Zhiyu Zhao [2], Weng-Hang Leong [3], Shining Chen [2], Xi Liu [1], Jingwei Fan [1,4], Xi Feng [1], Ren-Bao Liu [1,5,6,7], Yue Cui [2]\*, Quan Li [1,5,6]\*.*

1. Department of Physics, The Chinese University of Hong Kong, Shatin, New Territories, Hong Kong, China

2. Quantum Science Center of Guangdong-Hong Kong-Macao Greater Bay Area (Guangdong), Shenzhen 518045, China

3. Department of Engineering Science, Faculty of Innovation Engineering, Macau University of Science and Technology, Taipa, Macao 999078, China

4. School of Physics, Hefei University of Technology, Hefei, Anhui 230009, China

5. Centre for Quantum Coherence, The Chinese University of Hong Kong, Shatin, New Territories, Hong Kong, China

6. The State Key Laboratory of Quantum Information Technologies and Materials, The Chinese University of Hong Kong, Shatin, New Territories, Hong Kong, China

7. New Cornerstone Science Laboratory, The Chinese University of Hong Kong, Shatin, New Territories, Hong Kong, China

KEYWORDS

Nanodiamond, nitrogen–vacancy centers, single particle tracking, quantum sensing, multi-parameter sensing

ABSTRACT

Nitrogen-vacancy (NV) centers in nanodiamonds are excellent nanoscale sensors for measuring parameters such as temperature, magnetic field, and viscosity in complex fluidic environments, including living cells. However, the rapid motion of nanodiamonds in such dynamic systems imposes a significant challenge for continuous, real-time tracking and sensing measurements. Here, we present a fast single particle tracking (SPT) method featuring a tetrahedral detection geometry for time-efficient parallel fluorescence collection using four avalanche photodiodes (4-APDs), which eliminates the temporal latency of traditional sequential scanning. We demonstrate an improvement of about an order of magnitude in the temporal resolution and the upper limit of measurable diffusion coefficient compared to previously reported nanodiamond tracking methods based on single APD. The SPT is integrated with multi-parameter quantum sensing based on optically detected magnetic resonance (ODMR) of NV centers. The sensitivities of ODMR-based temperature and 3D rotation sensing are evaluated at different diffusion coefficients, which shows no significant degradation within our measurement range. We apply the system for thermorheology measurements in glycerol/water mixtures under thermal ramps. Additionally, we perform simultaneous translation and rotation tracking in live cells, revealing correlated translational and rotational dynamics. This approach advances multi-parameter nanoscale sensing for soft matter and biological applications, paving the way for real-time nanoscale sensing in highly dynamic fluidic environments.

# INTRODUCTION

Nanoscale sensing technologies have extended our ability to probe physical and chemical processes in complex environments ranging from soft matter systems[1,2] to living biological systems[3,4,5]. Among these platforms, nitrogen–vacancy (NV) centers in diamond, a point defect composed of a nitrogen atom adjacent to a lattice vacancy, are promising[6]. This stems from their long electronic spin coherence times under ambient conditions, enabling robust initialization, coherent manipulation, and optical readout of spin states without requiring cryogenic temperatures. The spin resonances of NV centers are highly sensitive to external perturbations that shift their energy levels, including magnetic fields[7], temperature[8], electric fields[9], and nuclear spins[10]. These properties are harnessed through optically detected magnetic resonance (ODMR), where laser excitation polarizes the spins into an excited state, microwave fields drive transitions between spin sublevels, and changes in fluorescence intensity reflect the resonance conditions. Nanodiamonds hosting NV centers exhibit robust physicochemical attributes, including chemical inertness, excellent photostability[11], high biocompatibility[12] and versatile surface functionalization[13–15], making them ideal for nanosensors such as thermometer[16–19], magnetometer[7,20,21], free radicals detectors[22], molecule sensor[23–25] and orientation sensor[2,4,5,26] in biological/soft matter systems.

An important characteristic of most biological/soft matter systems is that they exist in fluidic or quasi-fluidic environments. Consequently, single-particle tracking (SPT) becomes a cornerstone technique for investigating dynamical changes in biological systems[27,28], rheological properties of biomaterials[1,29,30], and properties of viscoelastic fluids[31–33]. Several nanodiamond SPT methods have been developed, based on wide-field or confocal techniques, demonstrating utility in biology application[4,34–37] and multi-parameters sensing[38]. However, wide-field methods[34,35,39] are commonly suited for two dimensional motion tracking and face a fundamental trade-off between axial sensitivity and temporal resolution while extending to three dimensions. Conventional tracking methods based on confocal setup are limited by the inherently low temporal resolution of multi-point sequential scanning around the target ND for determining its position[4,36,38], hindering the real-time tracking of fast-moving particles in 3D space which is important for dynamic measurements. Moreover, motion-induced fluorescence noise may degrade ODMR-based sensing sensitivity, highlighting the need for detailed assessment of nanodiamond sensing performance with tracking in dynamic environments.

In this work, we adopt a fast four-point-positioning tracking (4PPT) method using four avalanche photodiodes (4-APD) arranged in a tetrahedral geometry that simultaneously collects the photons from four slightly separated positions around an ND[40,41]. Compared to the conventional single-APD based tracking methods that rely on multiple sequential fluorescence acquisitions around the ND[4], the 4PPT method determines the 3D position of the target ND via a single fluorescence acquisition within one feedback cycle, and thus improve the tracking speed and temporal resolution. Besides, the simultaneous fluorescence collection by four APDs can quadruple the upper limit of the photon count rate imposed by a single APD, thereby enhancing the shot-noise-limited sensitivity. We demonstrate the method by tracking an ND on a cover glass with generated Brownian motion. One order of magnitude improvement in the measurement range of diffusion coefficient ($D<1.5\ \mu m^2/s$ *vs.* $D<0.16\ \mu m^2/s$ with 1-APD) and smaller tracking error have been achieved with the 4PPT method. ODMR measurement is integrated with single-particle tracking via temporally decoupled sampling timescales (40 μs for ODMR and 20 ms for tracking), allowing slow fluorescence noise to be averaged out in ODMR through the fast microwave sweeps. We investigate the impact of particle diffusion on the sensitivities of ODMR-based sensing. The sensitivities of temperature and 3D rotation sensing exhibit no significant degradation for diffusion coefficients up to the order of $10^{-1}$ $\mu m^2/s$. We further demonstrate the applications of the 4PPT platform for nanodiamond sensing in dynamic fluidic environments and living biological systems. Local thermo-rheology sensing was carried out in glycerol/water mixtures under thermal ramps, disclosing the correlation between temperature and particle (ND) motion. In simultaneous monitoring the translation and rotation of a ND sensor in live cells, the motion features suggest possible correlation between subcellular translational and rotational dynamics. The 4PPT based sensing system enriches the toolbox and brings new opportunities for studying dynamic processes in soft matter physics (e.g., glass transition rheology), as well as cell biology (e.g., cytoskeletal dynamics, membrane fluctuation and intracellular transport).

EXPERIMENTAL SECTION

**Setup**

Real-time SPT of NDs was performed on a home-built confocal system based on an Olympus IX-73 microscope frame (Figure 1a). A 532 nm laser, focused by a 60× oil immersion objective (Olympus, NA1.4), was used to excite the NV centers within the NDs. The confocal scanning was achieved using a 3-axis piezoelectric stage (Physik Instrumente, 517.3CD) to move the sample. To form a tetrahedral detection volume (shown in the inset of Fig.1a) around an ND sensor to determine its position, the fluorescence emitted from the sample were split into four collection pathways equally via three 50:50 cubic beamsplitter (BSW10, Thorlabs). The split fluorescence was then collected by an optical fiber in each of the pathway via a 10× collimating objective. The four optical fibers were coupled to 4-pack single-photon-counting avalanche photodiodes (Excelitas, SPCM-AQ4C).

The microwave (MW) was generated, amplified, and fed to a microwave antenna (a 25-µm copper wire) bonded on the cover glass substrate. The external magnetic field contains a static magnetic field in the *xy* plane applied by a pair of permanent magnets and a controllable magnetic field in the *z* direction by a magnetic coil mounted on the objective lens. The *z*-direction magnetic field was controlled by a direct-current (DC) source and an electronic switching circuit that enables reversible control of the field direction.

**Translational Tracking**

Translational tracking in this study was based on a parallel detection method utilizing multiple detectors to simultaneously determine the *x*, *y*, and *z* positions of particle. The four detection points (shown as red and blue dots P1-P4 in the inset of Figure 1a) forming the tetrahedral detection volume were positioned with P1 and P2 separated along the *y* axis and P3 and P4 separated along the *x* axis. The lateral (*x*, *y*) separations in the sample space were achieved by adjusting the reflection mirrors that direct the fluorescence into the optical fiber in the detection path. The vertical (*z*) separation was achieved by adjusting the distance between the collimating objective and the fiber pinhole, where we set the P1-P2 in the positive *z* direction and P3-P4 in the negative *z* direction. The experimental point spread functions (PSF) of the system are shown in SI Figure S1, estimated by the confocal scan images of a sub-diffraction-limited ND.

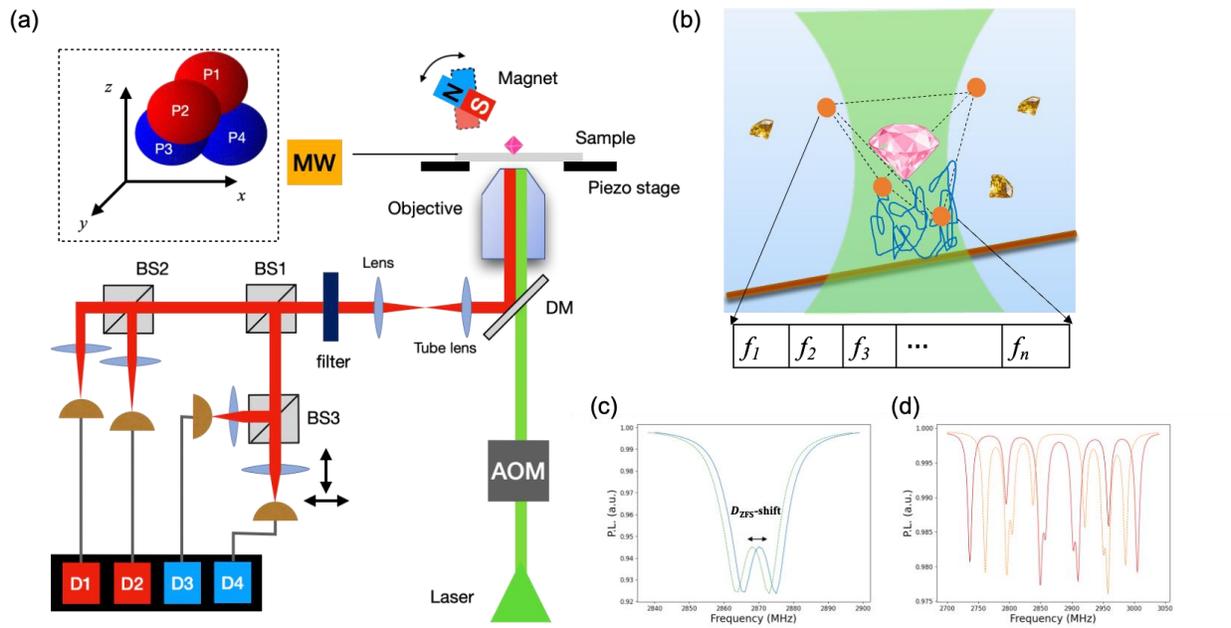

**Figure 1.** Single particle tracking for quantum sensing based on nanodiamond. (a) Schematic of the confocal microscope optics for 3D single particle tracking (SPT). The optical detection paths are configured to form a 3D tetrahedron-like detection volume in sample space. (b) Schematic of the ODMR-based sensing performed simultaneously with SPT. The emitted fluorescence from NV centers is collected at the four locations (orange dots) around the ND while the microwave (MW) is rapidly swept from $f_1$ to $f_n$. (c) Schematic of temperature sensing based on ND. The zero-field ODMR spectra at different temperature (blue line *vs* green line) shows a change in the shift of zero-field splitting $D_{ZFS}$. (d) Schematic of rotational sensing based on ND. The ODMR spectra of ND shows a change before (red line) and after a rotation (orange line) under an external magnetic field.

The tracking process started with an initial position around the target ND, with the fluorescence signals $S_1$-$S_8$ collected from the four positions (P1-P4) calculated as

$$\delta S_x = \frac{(S_1 + S_4) - (S_2 + S_3)}{\max(S_1 + S_4, S_2 + S_3)},$$

$$\delta S_y = \frac{(S_1 + S_3) - (S_2 + S_4)}{\max(S_1 + S_3, S_2 + S_4)},$$

$$\delta S_z = \frac{(S_1 + S_2) - (S_3 + S_4)}{\max(S_1 + S_2, S_3 + S_4)}.$$

The coordinate system here is rotated by 45° around the *z*-axis relative to that given in Figure 1a to ensure that the signals from the four detectors were fully incorporated into the calculation of the *x*- and *y*-axis positions. The three error signals serve as the inputs for the proportional integral differential (PID) controllers to determine which direction the stage should be moved to reposition the ND closer to the center of the tetrahedral detection volume. The three PID output voltages were fed back to drive the piezoelectric stage to its new target position. The stage movement could thus follow the target ND and record the 3D trajectory. ND fluorescence was sampled for a dwell time of 20 ms in each feedback cycle. The cycle time was approximately an order of magnitude shorter than that of the multi-point (six-point or eight-point) scanning tracking method based on a single APD[4,42].

**ND-based Sensing by ODMR**

*Simultaneous ODMR and Single Particle Tracking.* In this work, ND based sensing was carried out by continuous-wave (CW) ODMR. As presented in Figure 1b, the ODMR detection and translation tracking was encoded in the photon counts upon MW frequency sweeping and position adjustment simultaneously. The MW frequency was swept through $f_1, f_2, …,$ and $f_n$, with a dwell time of 40 μs for each frequency (much smaller than the 20 ms dwell time of translation tracking). For each frequency, the fluorescence from four channels collected by 4-pack APD were counted by a data acquisition board system (National Instruments, PCIe-6363) and summed. To realize a sufficient signal-to-noise ratio (SNR), the frequency sweeping cycle ($f_1$ to $f_n$) was repeated 80 times to obtain a line of the full-spectrum ODMR. This acquisition time defines the temporal resolution of the CW-ODMR-based sensing, which is typically on the order of hundreds of milliseconds (depending on the number of frequencies).

*Temperature Sensing.* For temperature sensing based on ND, we performed ODMR without external magnetic field. With variation in temperature, the zero-field splitting $D_{ZFS}$ exhibits a detectable shift, as schematically depicted in Figure 1c. We followed standard protocol[8] to full-spectrum ODMR containing 31-points sampling MW frequency was acquired with a total time of ~0.1 s. The fluorescence signal was normalized to the counts collected at an off-resonance MW frequency based on a target integration time. The normalized ODMR spectra was fitted with double-peak Lorentzian function to obtain the zero-field splitting $D_{ZFS}$, which has a dependence on environment parameters such as temperature (see details in SI Section S1).

*Rotation Sensing.* For rotation sensing based on ND, the ODMR spectra was collected under external magnetic field. as schematically shown in Figure 1d, ODMR spectra changes under rotation. To eliminate ambiguity in determining the ND orientation and to obtain full 3D rotation information, ODMR spectrum should be collected under two different magnetic fields[2,4]. We followed a standard protocol reported earlier[4]. Briefly, for the dual-magnetic-field experiment, ODMR spectra were recorded sequentially: following the acquisition under the first magnetic field, the magnetic field was switched and a second ODMR spectrum was collected. The total acquisition time for one cycle, including field switching and dual-spectra collection, was approximately 1 s. By simultaneously fitting both spectra to a multi-peak model accounting for all four NV crystallographic orientations, we obtained the Euler angles (α, β, γ) describing the ND's 3D orientation (details in SI Section S1). For experiments demanding higher temporal resolution, we adopted the single-magnetic-field method to reduce acquisition time but sacrificing the full 3D rotational data. The ODMR spectra was acquired at a single magnetic field, with an integration time of approximately 0.5 s.

**Nanodiamonds**

The fluorescent NDs (with particle size in the range of hundreds of nm) used in the experiments were purchased from FND Biotech (Taiwan). These NDs contain over 1000 NV centres each on average. For tracking experiments in cell cytoplasm, NDs were coated with BSA (bovine serum albumin, Sigma-Aldrich) to avoid particle aggregation in culture medium. For tracking experiments on cell membrane, NDs were coated with fibronectin (Sigma-Aldrich) to facilitate specific binding to the membrane. The 100 μL ND solution (1 mg/mL) were first thoroughly sonicated for 10 min, then mixed with 100 μL BSA (5 mg/mL) or 100 μL fibronectin (0.1 mg/mL) and 800 μL DI water. The suspension was gently shaken overnight at room temperature. Afterwards, unbound proteins were removed by centrifugation (20°C, 10000 rpm, 5 min) for three times. The pellet was washed extensively with water and finally suspended in 1 mL phosphate buffered saline (PBS). The dispersibility of bare and protein-coated NDs in DI water and PBS were assessed by measuring their hydrodynamic particle size using dynamic light scattering (DLS) (see SI Figure S2).

**Sample Preparation**

*Stationary NDs on Cover Glass.* ND solution was diluted in ethanol to a concentration of 1 μg/mL and then ultrasonicated for 5 minutes to avoid potential aggregation. Subsequently, 200 μL of the ND dispersion was loaded into an atomizer (Paasche, HS202S) using a micropipette. The atomizer's pressure was optimized to generate a monodisperse aerosol for uniform deposition of NDs across the cover glass. The cover glass was fixed to a printed circuit board (PCB) with a microwave antenna subsequently soldered onto the sample surface.

*NDs in Glycerol/water Solution.* To make a suspension of nanodiamonds in 70% glycerol/water solution (weight ratio), 5 μL of 1 mg/mL ND solution was diluted in 295 μL DI water and then added to 0.7 g of glycerol. The suspension was thoroughly mixed before degassing under vacuum and then shaken for at least 4 hours. The final concentration of ND in the solution was 5 μg/g.

A pre-cut double-sided tape (with dimensions of 200 μm in height and 1 mm in diameter) was attached to a cover glass pre-mounted on a PCB integrated with a microwave antenna. Subsequently, 100 μL of the pre-prepared glycerol/water solution with NDs was pipetted into the reservoir formed by the double-sided tape. Thereafter, another cover glass was placed on top of the tape to seal the reservoir. To prevent potential solution leakage, ultraviolet (UV) curing glue was uniformly applied around the edge of the upper cover glass.

*Culture of Cells and Incubation with NDs.* HeLa and MCF-7 cells were cultured in growth medium [Dulbecco's modified Eagle's medium (DMEM, Gibco) supplemented with 10% fetal bovine serum (FBS, Gibco) and 1% penicillin-streptomycin (Gibco)] at 37°C with the $CO_2$ level (5%) and humidity controlled.

A confocal dish was prepared on a PCB with microwave antenna bonded to the surface of the cover glass. The substrate was treated with plasma to enhance cell adhesion. Then the cells were seeded in the dish and cultured with DMEM for 24 hours before introducing NDs. For experiment performed on cell membrane, The fibronectin coated NDs were dispersed in DMEM and incubated with MCF-7 cell for 1 hour at 4°C to allow attachment while inhibiting cell endocytosis before performing the 6D tracking (at 37°C). For tracking experiments in cytoplasm, the BSA coated NDs were dispersed in DMEM-FBS medium and incubated with HeLa cells for 6 hours at 37°C in the incubator. Excess NDs were removed by washing the cell three times with PBS and fresh DMEM-FBS medium was then added. The dish was placed into a live-cell workstation (Tokaihit, STXG-WSKMX-E) mounted on the microscope. The system maintained an environment at 37°C with 5% $CO_2$ during the sensing experiments.

**Heating Device with Temperature Control**

A custom-built water bath heating device with water reservoir and heating ceramic was adopted for controlling and monitoring temperature during the glycerol/water experiments. A Pt 1000 resistance temperature detector (RTD) is placed on the top of the upper cover glass with the sensor covering the central region across the microwave antenna. The heating ceramic and RTD were connected to a programmable temperature PID controller (TCM-M207, YeXian).

**Oscillatory Rheology Measurement**

The viscosity of 70% glycerol/water solution was measured by a rheometer (MCR 302e, Anton-Paar) equipped with a cone-plate measuring system (CP25 with a diameter of 25 mm). Rheology measurement with a temperature sweep from 20 to 60°C was performed with a sampling rate of 0.5 min/points and a shear rate of 50 $s^{-1}$.

## RESULTS AND DISCUSSION

### Evaluation of the Translational Tracking Performance

We evaluated the performance of the 4PPT method using an ND on cover glass with generated translational motion (serving as the reference trajectory), which was designed to mimics Brownian motion by moving the piezo stage on which the ND sample was mounted (see details in SI section S3). Figure 2a compares the tracking trajectory detected by the 4PPT tracking method and the reference one in the $x$, $y$, and $z$ axes with a given diffusion coefficient of $D = 0.5$ $\mu m^2/s$. The tracking errors ($\delta$) in each axis are plotted in the respective bottom panel. As a comparison, similar measurements were conducted using a 1-APD translation tracking system[4] at a given $D = 0.05$ $\mu m^2/s$. Although there is one order of magnitude difference in the set diffusivity, a similar error range was found in the tracking results of the 4PPT and the 1-APD system. The larger error observed in the $z$-tracking (than that of the $x$- and $y$-tracking) is ascribed to the inferior optical resolution in $z$ directions (SI Figure S1). We further compared the 4PPT and 1-APD methods at different given diffusion coefficients, as shown in Figure 1c. Both systems show a good match between the measured diffusivity and the input reference (Figure 1c). However, the 4PPT system shows a much larger measurement range with upper limit of $D = 1.5$ $\mu m^2/s$ with smaller tracking error within the measurement range, compared to those in the 1-APD system (Figure 1c, SI Figure S3). The larger upper bound and reduced dynamic error of the 4PPT system are attributed to its faster feedback cycle enabled by the parallel detection architecture.

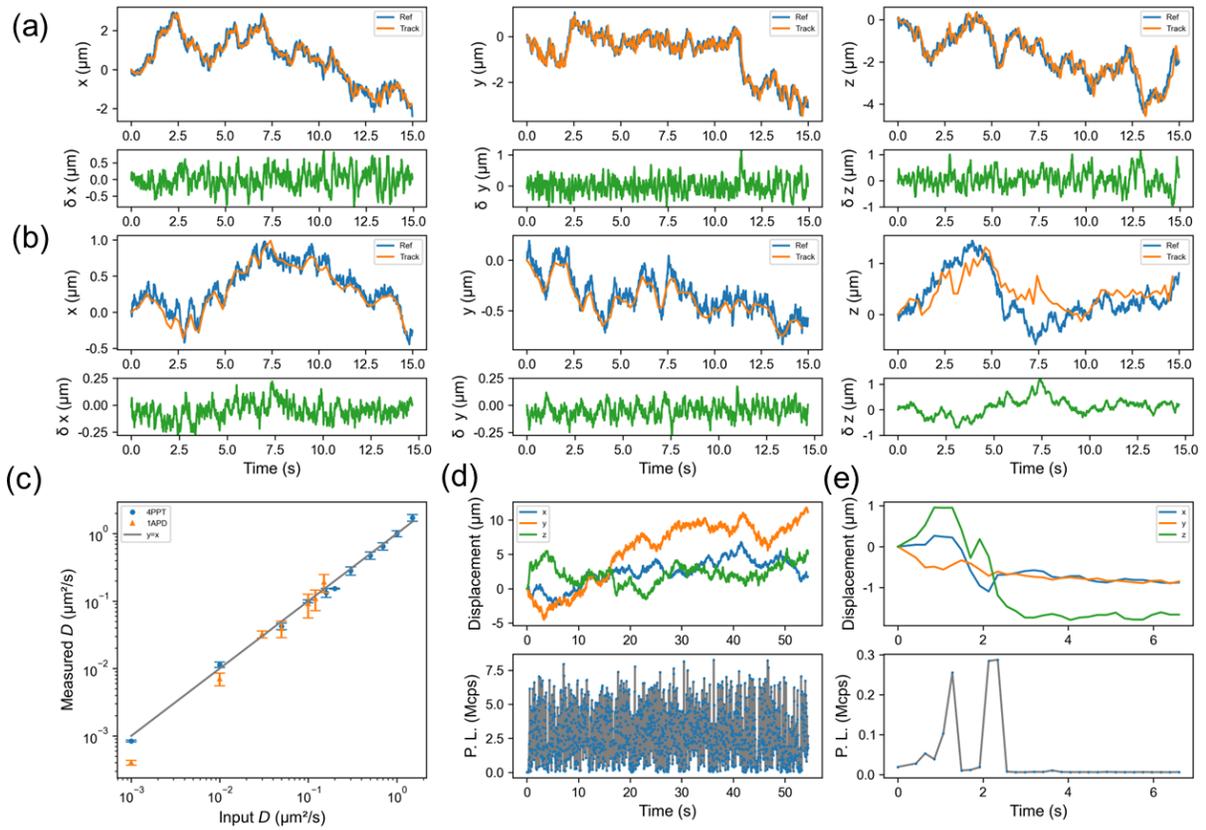

**Figure 2.** Evaluation of the translational tracking performance. (a) The comparison between the tracking trajectory (orange lines) and the reference particle trajectory (blue lines) measured by the 4PPT system with a diffusion coefficient of $D = 0.5\ \mu m^2/s$. The corresponding difference between the two trajectories is shown in the lower panel (green lines). (b) The same comparison performed on the 1-APD based tracking system with $D = 0.05\ \mu m^2/s$. (c) Comparison of the input and the measured diffusion coefficients obtained (D) from the 1-APD (orange dots) and 4PPT systems (blue dots). The measured D are in good agreement with the input $D$ (grey line). (d, e) Comparison of the tracking performance in 50 wt% glycerol/water solution ($D \sim 0.4\ \mu m^2/s$)

based on (d) the 4PPT system and (e) the 1-APD system. The upper panels show the 3D tracking trajectories and the lower panels show the photon counts *vs* time during tracking.

We then compared the 4PPT and 1-APD based tracking method in real systems, that is, tracking NDs in a 50 wt% glycerol/water solution (with a diffusion coefficient of $D\sim0.4~\mu m^2/s$ as shown in Figure S4). The 4PPT based system maintained a stable tracking of an ND in the 60 s tracking duration (Figure 2d), while the 1-APD based system lost the target (ND) within a couple of seconds (Figure 2e).

**Evaluation of the Sensing Performance of ND during Real-Time Tracking**

We first evaluated the sensing performance of ND during its real-time tracking by carrying out ODMR measurement of an ND on cover glass undergoing a generated reference Brownian motion. Figure 3a show the 3D trajectory of an ND measured at a diffusion coefficient of $D = 0.8~\mu m^2/s$. In the absence of external magnetic field, a typical ODMR spectrum simultaneously taken from this ND is shown in Figure 3b. The ODMR signal shows no significant difference compared to that obtained under stationary conditions (shown in SI Figure S5). We then evaluated the influence of dynamic motion and tracking of ND on the sensing performance by analyzing the measurement sensitivity at different diffusion coefficients. The sensitivity was estimated by plotting the standard deviation of the $D_{ZFS}$-shift as a function of the inversed square root of the data acquisition time (SI Figure S6). As plotted in Figure 3c, the estimated sensitivity shows no significant dependence on the diffusion coefficient within the measurement range.

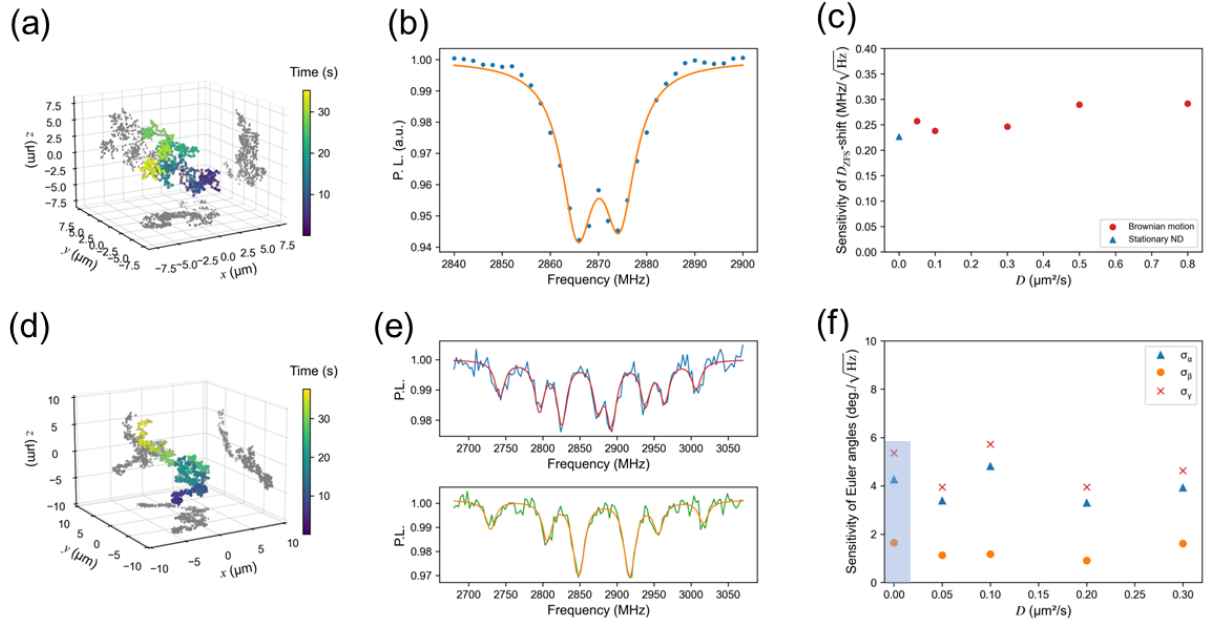

**Figure 3.** Evaluation of the temperature and rotation sensing performance of ND during real-time tracking based on the 4PPT system. (a) The trajectory and (b) the zero-field ODMR spectrum (blue dots) with fitting (orange line) of an ND undergoing a generated Brownian motion with $D = 0.8\ \mu m^2/s$, tracked by the 4PPT based system. (c) The measured $D_{ZFS}$-shift sensitivity as a function of the diffusion coefficient of Brownian motion (red dots). The blue triangle represents the sensitivity obtained on the stationary ND on cover glass without motion. (d) The trajectory and (e) the ODMR spectra (blue and green lines) with fitting (red and orange lines) under two magnetic fields of an ND undergoing a generated Brownian motion with $D = 0.3\ \mu m^2/s$. (c) The sensitivities of the three Euler angles (α, β, γ) as functions of the diffusion coefficient of Brownian motion. The blue shaded area shows the data extracted from the stationary ND.

For rotational sensing, we applied a dual-magnetic-field protocol with two sequential ODMR spectra acquired under two different magnetic fields (~50 G). Figure 3d and 3e show the 3D trajectory and ODMR spectra under two magnetic fields of an ND measured under a diffusion coefficient of $D = 0.3\ \mu m^2/s$. The sensitivities of the three Euler angles (α, β, γ) describing the ND rotation were extracted (SI Figure S7) and plotted as a function of $D$ (Figure 3f). In the measurement range, the rotation sensitivity shows no significant dependence on the diffusion coefficient.

**Nanoscale Thermorheology Sensing in Glycerol/Water Mixtures**

The proof-of-the-concept demonstration of nanoscale thermorheology was carried out by tracking both the temperature measured by ND and its 3D trajectory. During thermal ramps of 70 wt% glycerol/water solution (in which the diffusion of ND exceeds the tracking limit of the 1-APD system), the ODMR and the 3D trajectory of a single ND were monitored in real time when the temperature was raised from ambient to 60 °C (measure by RTD). A down-shift of the zero-field splitting was observed in the recorded ODMR spectra (see SI Figure S8). As shown in Figure 4a, the magnitude of the $D_{ZFS}$ shift (orange dots) increases with time upon heating (at 200 s), whose profile agreed well with the temperature evolution recorded by the reference RTD on the cover glass (the grey curve). The in-situ, local rheological changes of the solution during the heating process are reflected by the instantaneous velocity of the ND, which is derived from its 3D trajectory data. The ND velocity shows an overall increase with heating (Figure 4b), along with fluctuations likely contributed from stochastic Brownian motion and thermal convection within the solution.

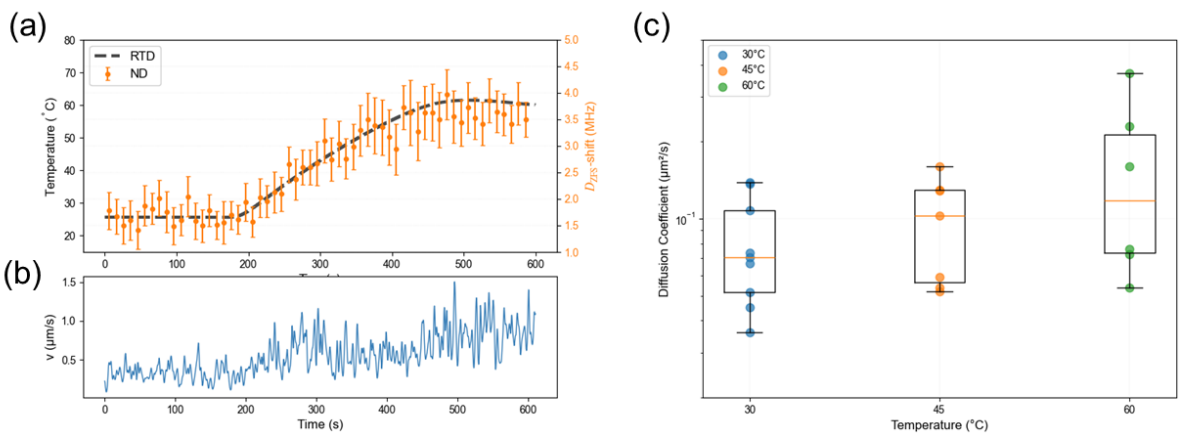

**Figure 4.** Simultaneous nanothermometry and nanorheometry in glycerol/water solution. (a) The temperature measured by RTD (grey dashed curve) and the $D_{ZFS}$-shift extracted from ODMR (orange dots) of an ND under a continuous heating process. (b) The instant velocity obtained from the tracking trajectory of the ND during heating. (c) Box plots comparing the diffusion coefficients of multiple NDs measured at three different temperatures (measured by RTD). The central orange line denotes the median value, the box represents the interquartile range, and the whiskers extending from the box indicate the range of the data.

The rheology property of the solution was investigated by analyzing the 3D trajectories (see SI Figure S9) of multiple NDs at thermal equilibrium (T = 30, 45, and 60°C). The diffusion coefficients ($D$) of the NDs were extracted from their respective MSD (see SI Section 2) analysis. As shown in Figure 4c, a spread in the diffusion coefficient at any specific temperature is ascribed to the size dispersion of the ND particles, as the diffusion coefficient depends on temperature, particle size and the solution viscosity[43]. Despite the dispersion, an increase in the diffusion coefficient was found with temperature in the box plot of $D$. Taking the temperature-dependent solution viscosity measured by bulk-rheometry (SI Figure S10a), the sizes of the NDs were extracted from the Stokes-Einstein relation using the measured *D* from ND tracking (see details in SI Section S2). The calculated range of particle diameter (~100-800 nm, see SI Figure S10b) is consistent with that estimated from the dynamic light scattering data (SI Figure S10c).

**Simultaneous Translation and Rotation Tracking in Live Cell system**

Simultaneous rotation and translation tracking of ND using the 4PPT system was first tested by monitoring the 6D motion of a ND on a live cell plasma membrane, on which the motion of ND is rather slow. Figure 5a shows the fluorescence image of an MCF-7 cell with an ND anchored on its plasma membrane. The translational trajectories of the ND over 900 seconds are plotted in Figure 5b, exhibiting a slow, confined translational motion within a volume of approximately 1 $\mu m^3$. The confined translational motion is consistent with the literature reports (Ref [4,44]) and is generally understood as resulted from the specific binding of the fibronectin-coated NDs to the immobilized integrin receptors that are anchored to the actin network beneath of the plasma membrane. Figure 5c shows the time-dependent ODMR simultaneously recorded with translation tracking at two different magnetic fields. The rotation of the ND, manifested by the large fluctuations in the ODMR resonance frequencies, is associated with the dynamic deformations of the plasma membrane surrounding the ND[4]. The instantaneous angular speed can be deduced from the rotation trajectory obtained by fitting the two ODMR spectra (see SI Section S1). Figure 5d shows the extracted translational and angular speed of the ND in real time. The results are consistent with those obtained using a 1-APD tracking system reported earlier as in Ref [4].

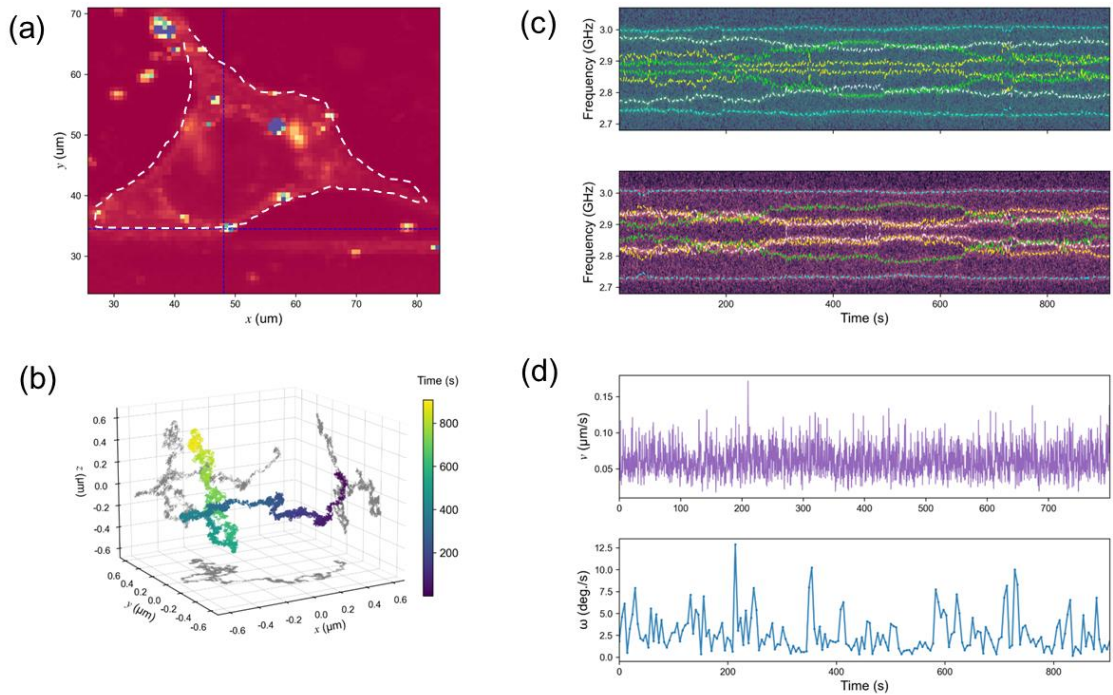

**Figure 5.** Translation and rotation tracking of a nanodiamond on the membrane of a live MCF-7 cell. (a) *xy*-cross of the confocal fluorescence image of the MCF-7 cell (enclosed by the white dashed lines). An ND attached on the plasma membrane is indicated by the intersection of two blue dashed lines. (b) Translational trajectory of the ND on the membrane (time shown in color scale), together with 2D projections to the $x-y$, $y-z$, and $x-z$ planes (shown in grey dots). (c) Time-dependent ODMR spectra with fitting of the ND under two magnetic fields. The fitting results (dashed lines) are overlapped on the raw date. (d) Translational (the upper panel) and angular (the lower panel) speed of the ND, calculated from the translation trajectory and the fitting results of the ODMR spectra, respectively.

While the translational motions of nanoparticles anchored on the plasma membrane are slow ($v \sim 0.1$ μm/s) in general, these nanoparticles can speed up once inside the cells, depending on their local environment. We therefore monitor the motion of an ND in the cytoplasm of a live HeLa cell. Figure 6a presents the fluorescence image of a HeLa cell with an internalized ND marked by the crossing point of the blue dashed lines. The real-time 3D tracking trajectories of the ND are shown in Figure 6b, in which one can find the individual $x$, $y$, and $z$ trajectories. The corresponding instantaneous translation velocity extracted from the trajectories is plotted as a function of time in Figure 6c, in which one can divide the translational motion into two stages with the first one (0-480 s, blue shadow) featured by frequently appearing high instant velocity, and the second one (>480 s, purple shadow) characterized by low instant velocity.

Further analysis of the ND's translation in the two stages was carried out by quantitatively analyzing the ND translation dynamics. By dividing the translational trajectories into 10-second segments throughout the entire tracking process, here we plotted the mean square displacements (MSDs) of each segment as a function of $\tau$ (Figure 6d), where $\tau$ is the time interval (details can be found in SI Section S2). Fitting of the MSD profiles with the power-law scaling relation MSD $\propto \tau^\alpha$ gave $\alpha$, which is the anomalous diffusion exponent. In Figure 6d, the data extracted from the two stages are presented in two different colors (blue and purple). The translational behaviors of ND in the first stage (blue lines) are dominated by diffusive Brownian motion ($\alpha \sim 1$) and super-diffusive transport ($\alpha > 1$). The MSD of a representative segment (445-455 s, indicated with orange arrow in Figure 6b) with directional motion is plotted in Figure 6d. The segment shows an anomalous exponent $\alpha$ of 1.6, indicating an intracellular active trafficking processes[38,45,46]. As a comparison, the motions in the second stage (purple lines) are predominantly confined ($\alpha < 1$). The MSD value in the second stage is on the order of $10^{-3}$ to $10^{-2}$ μm$^2$/s, approximately one order of magnitude lower than that in the first stage.

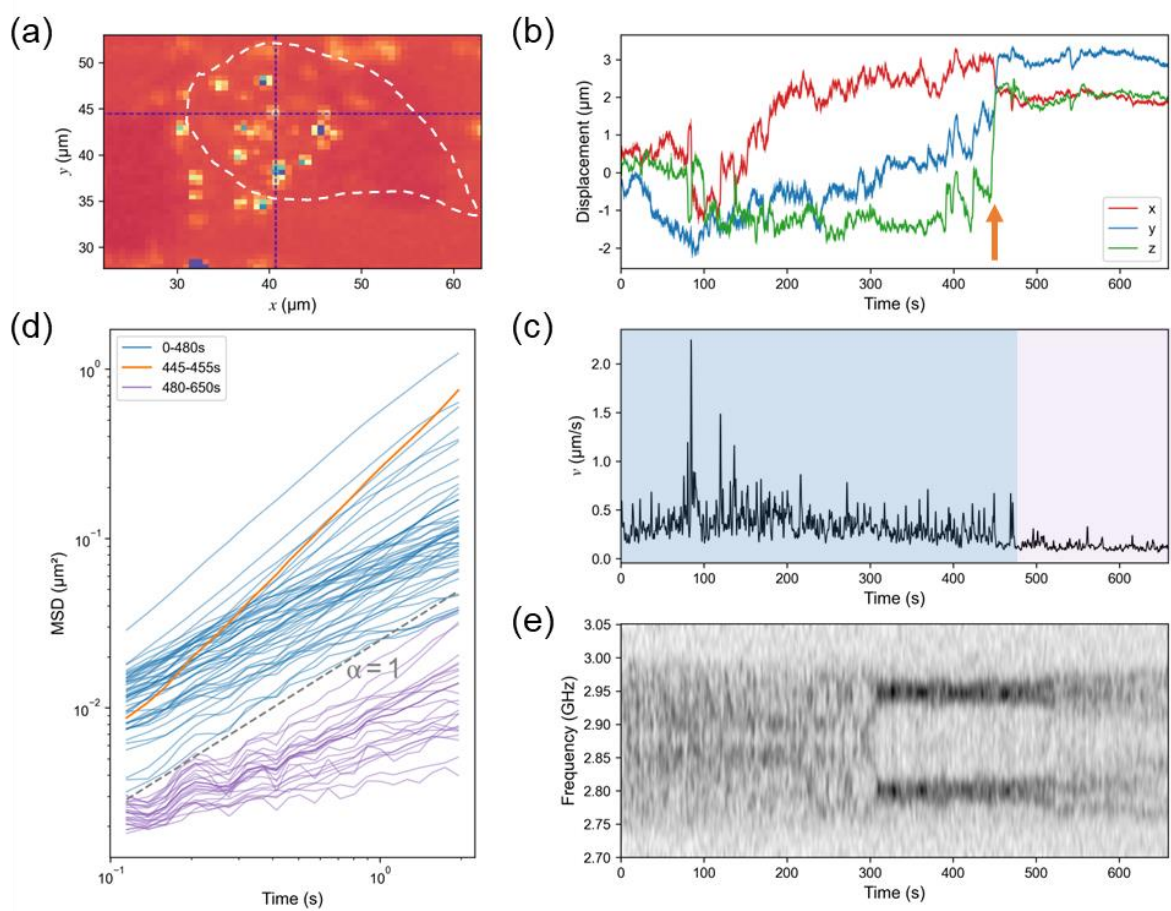

**Figure 6.** Translation and rotation tracking of a nanodiamond in the cytoplasm of a live Hela cell. (a) *xy*-cross of the confocal fluorescence image of the Hela cell (enclosed by the white dashed lines). An ND in the cytoplasm is indicated by the intersection of two blue dash lines. (b) Real-time displacements of the ND in *x*, *y*, and *z* directions. The orange arrow indicates a time interval (445-455 s) with directional motion. (c) Instant translational speed of the ND extracted from the trajectory. (d) The mean square displacement (MSD) of the ND translational motion during the tracking process, with the data colored according to stage: blue for 0-480 s, and purple for 480-650 s, orange for 445-455 s. The grey dash line shows a curve with $\alpha = 1$. (e) Time-dependent ODMR spectrum of the ND with the contrast of ODMR encoded in the greyscale.

The translation characteristics of the ND in the first stage, that is, fast random motions accompanied by occasional directed transport segments, suggest a Brownian-motion-like behavior of the NDs in the crowded, heterogeneous intracellular environments[38,45,47,48]. The directed-motion segments, such as the fast bidirectional transport of ND with a maximum speed exceeding 2 μm/s (e.g., at ~90 s), may be related to the cargo motion at track-switching or transport processes in living cell[49]. In the following stage, the rather slow translation of ND, together with the restricted motion characteristics suggest that the ND may become trapped by certain subcellular structures (such as the F-actin-rich cellular cortex[50,51]).

Rotation tracking in the cytoplasm is very challenging as the rotation angular velocity can be extremely large, while the existing full 3D rotation tracking is slow. We therefore adopted the single-magnetic-field OMDR measurement to record the rotation behavior of the ND (see Experimental Section). The real-time ODMR, simultaneously recorded with translation tracking is shown in Figure 6e. In the first stage (0-480 s) when the translation is dominated by Brownian motion with fast directional sectors, two different rotation characteristics were observed for the ND. In the very first 300 s, the ND exhibited rapid random rotation resulting in fast changing resonance frequency (barely resolvable) in the recorded ODMR. However, from 300-480 s, the real-time ODMR spectrum stabilized with little change in the resonance frequency, suggesting that the rotation was either quenched or proceeded around the applied magnetic field[2,4]. This behavior (confined rotation with directional translation) matches the pattern of microtubule-associated trafficking processes[42,46], where the orientation of the nanoparticle did not significantly change over the transport. In the next stage characterized by a slow and confined translational motion (~480 s to the end), the clearly resolvable resonance frequency in the recorded ODMR (Figure 6d) suggested a rotation scenario similar to that recorded on the plasma membrane in Figure 5c.

It is worth noting that the maximum translational tracking speed of a single APD system is ~1 µm/s, making it difficult to track the 3D motion of fast nanosensors (ND in the specific case). Here we show that the 4PPT-based tracking system brings in one order of magnitude improvement in the translational tracking, and at the same time allowing ODMR readout from fast-moving nanodiamond sensors without losing measurement sensitivity. With further development of parallel cellular structure staining and fast rotation tracking method, the 4PPT tracking system provides a basic platform for studying intriguing transport dynamics in live cells. The dual-modal (translational and rotational) rheological measurement provided a more comprehensive investigation of intracellular dynamics, facilitating the association of sensor motion behaviors with subcellular structures or processes.

CONCLUSIONS

This work establishes a high-speed single particle tracking platform that is compatible with NV- based quantum sensing, with instantaneous tracking speed reaching several µm/s and diffusion coefficients up to 1.5 µm$^2$/s. We show that standard sensing measurement (CW-ODMR) can be carried out along with fast particle tracking without sacrificing the sensitivity. The platform's multi-parameter sensing capability is further validated via thermorheology measurement in glycerol/water mixtures and 6D tracking in live cell systems. The platform is particularly suitable for studying dynamics in complex fluidic environments and live cell systems, which have typical diffusion coefficients ranging from ~0.01 to 1 µm$^2$/s and instant speed exceeding 1 µm/s during non-equilibrium transport processes[45,54–58]. The 4PPT system thus provides a valuable tool for nanoscale multi-parameter sensing, with great potential for wide-range applications in soft matter physics, biophysics, and cell biology.

**Supporting Information**.

The following files are available free of charge. Measured PSFs of the four detection volumes, DLS results, comparison of tracking error, the mean square displacement (MSD) of ND, comparison of ODMR spectra, standard deviation of $D_{\text{ZFS}}$-shift and Euler angle, the dependence between ZFS frequency and temperature, typical real-time trajectory of the NDs under different temperature, bulk viscosity measured by rheometer and size of NDs, temperature and rotation analysis based on ODMR, translational and rheology analysis for Brownian motion, simulation of Brownian motion trajectories (PDF)

AUTHOR INFORMATION

**Corresponding Author**

Yue Cui − *Quantum Science Center of Guangdong-Hong Kong-Macao Greater Bay Area (Guangdong), Shenzhen 518045, China;* orcid.org/0000-0002-3561-0453; Email: *cuiyue@quantumsc.cn*

Quan Li − *Department of Physics, The Chinese University of Hong Kong, Shatin, New Territories, Hong Kong 999077, China; Centre for Quantum Coherence and State Key Laboratory of Quantum Information Technologies and Materials, The Chinese University of Hong Kong, Shatin, New Territories, Hong Kong 999077, China;* orcid.org/ 0000-0001-8563-1802; Email: *liquan@cuhk.edu.hk*

**Author Contributions**

[‡]G.Z. and M.Z.A. contributed equally. Q.L. conceived the idea and supervised the project. G.Z., Y.C., M.Z.A., and Q. L. designed the experiments. G.Z. and M.Z.A. constructed the set up with

the contribution from Y.C., X.F., J. W. F., and X.L.. G.Z. performed the experiments under the supervision of Y.C.. G.Z. and S.N.C. prepared the samples. G.Z., Y.C., Z. Y. Z., W.-H.L., R.-B.L., and Q.L. analyzed the data. G.Z. Y.C., M.Z.A., R.-B.L., and Q.L. wrote the paper, and all authors commented on the manuscript.

**Funding Sources**

This work was supported by RGC/CRF of HKSAR: no. C4004-23G, RGC/GRF of HKSAR: no. 14301722, NSFC: no. 12404581, Guangdong Provincial Quantum Science Strategic Initiative under grants: no. GDZX2305007 and GDZX2405001, the Start-up Funding for the Quantum Science Center of Guangdong-Hong Kong-Macao Greater Bay Area: no. QD2405001, Innovation Program for Quantum Science and Technology: no. 2023ZD0300600.

**Notes**

The authors declare no competing financial interest